# Electron pockets in the Fermi surface of hole-doped high-$T_c$ superconductors

David LeBoeuf [1], Nicolas Doiron-Leyraud [1], Julien Levallois [2], R. Daou [1],

J.-B. Bonnemaison [1], N.E. Hussey [3], L. Balicas [4], B.J. Ramshaw [5], Ruixing Liang [5,6],

D.A. Bonn [5,6], W.N. Hardy [5,6], S. Adachi [7], Cyril Proust [2] &  Louis Taillefer [1,6]

1 Département de physique and RQMP, Université de Sherbrooke, Sherbrooke J1K
2R1, Canada

2 Laboratoire National des Champs Magnétiques Pulsés (LNCMP), UMR CNRS-UPS-
INSA 5147, Toulouse 31400, France

3 H.H. Wills Physics Laboratory, University of Bristol, Bristol BS8 1TL, UK

4 National High Magnetic Field Laboratory, Florida State University, Tallahassee,
Florida 32306, USA

5 Department of Physics and Astronomy, University of British Columbia, Vancouver
V6T 1Z4, Canada

6 Canadian Institute for Advanced Research, Toronto M5G 1Z8, Canada

7 Superconductivity Research Laboratory, International Superconductivity Technology
Center, Shinonome 1-10-13, Koto-ku, Tokyo 135-0062, Japan

High-temperature superconductivity occurs as copper oxides are chemically tuned to have a carrier concentration intermediate between their metallic state at high doping and their insulating state at zero doping. The underlying evolution of the electron system in the absence of superconductivity is still unclear and a question of central importance is whether it involves any intermediate phase with broken symmetry[1]. The Fermi surface of underdoped $YBa_2Cu_3O_y$ and $YBa_2Cu_4O_8$ was recently shown to include small pockets[2,3,4] in contrast with the large cylinder characteristic of the overdoped regime[5], pointing to a topological change in the Fermi surface. Here we report the observation of a negative Hall resistance in the magnetic field-induced normal state of $YBa_2Cu_3O_y$ and $YBa_2Cu_4O_8$, which reveals that these pockets are electron-like. We propose that electron pockets arise most likely from a reconstruction of the Fermi surface caused by the onset of a density-wave phase, as is thought to occur in the electron-doped materials near the onset of antiferromagnetic order[6,7]. Comparison with materials of the $La_2CuO_4$ family that exhibit spin/charge density-wave order[8,9,10,11] suggests that a Fermi surface reconstruction also occurs in those materials, pointing to a generic property of high-$T_c$ superconductors.

The Hall effect is a powerful probe of the Fermi surface of a metal because of its sensitivity to the sign of charge carriers, able to distinguish between electrons and holes. In addition, the Hall effect has been the prime transport signature of density-wave order in cuprates such as $La_{2-y-x}Nd_ySr_xCuO_4$ (Nd-LSCO) (ref. 10) and $La_{2-x}Ba_xCuO_4$ (LBCO) (ref. 11). The Hall resistance $R_{xy}$ of LBCO is reproduced in Fig. 1, where it is seen to drop precipitously below a temperature $T_{DW}$ which coincides with the well-established onset of spin/charge density-wave order in this material[8]. The drop leads to a change of sign in $R_{xy}$, pointing to a reconstruction of the Fermi surface from purely hole-like above $T_{DW}$ to a combination of electron-like and hole-like sheets below. The fact that our high-field measurement of $R_{xy}$ in $YBa_2Cu_3O_y$, a cuprate material with a different structure and considerably higher purity, cation order and maximal $T_c$, exhibits a similar behaviour of $R_{xy}$, as shown in Fig. 1, raises the possibility that Fermi surface reconstruction may be a generic phenomenon in cuprates, and hence likely to be essential for a full understanding of high-temperature superconductors.

The Hall resistance $R_{xy}$ was measured in two closely related underdoped cuprates of the YBCO family: YBa$_2$Cu$_3$O$_y$ (Y123), with $y = 6.51$ and $y = 6.67$, and YBa$_2$Cu$_4$O$_8$ (Y124). The Y123 samples have a high degree of oxygen order, with ortho-II and ortho-VIII superstructure, respectively. The Y124 is stoichiometric, with intrinsic oxygen order. With a $T_c$ of 57.5 K, 66.0 K and 80 K, respectively, the three samples have a hole doping per planar copper atom of $p = 0.10$, 0.12 and 0.14, respectively, *i.e.*, they all fall in the underdoped region of the doping phase diagram (*i.e.* $p < 0.16$). (Sample characteristics are given in the Methods Summary.) The current was applied along the $a$-axis of the orthorhombic structure ($J \parallel x \parallel a$), *i.e.*, perpendicular to the CuO chains, in magnetic fields applied normal to the CuO$_2$ planes ($B \parallel z \parallel c$). (Details of the measurements are given in the Methods Summary.) The Hall coefficient $R_H \equiv t\, R_{xy} / B$, where $t$ is the sample thickness, is displayed as a function of magnetic field in Figs. 2a to 2c and as a function of temperature in the Supplementary Information (Figs. S1a to S1c).

Our central finding is that all three materials have a negative Hall coefficient in the normal state at low temperature. This is displayed in Fig. 3, where a plot of $R_H$ vs $T$ at the highest field reveals a change of sign from $R_H > 0$ above $T = T_0$ to $R_H < 0$ below, with $T_0 = 30$, 70 and 30 K, for II, VIII and Y124, respectively, with ± 2 K uncertainty. A very similar sign change was reported by Harris *et al.*[12] in Y123 samples with $T_c = 62$-64 K. Because their measurements were limited to moderate fields (below 24 T), these authors attributed the negative $R_{xy}$ to a negative contribution to the Hall conductivity $\sigma_{xy}$ coming from vortices (flux flow). By going to much higher fields, we can now rule out this interpretation, as discussed in detail in the Supplementary Information, where the negative $R_H$ is shown to be unambiguously a property of the normal state, the consequence of a drop in $R_H(T)$ which starts below a field-independent temperature $T_{max}$. The value of $T_{max}$ at the three doping levels studied here is 50, 105 and 60 K, for II, VIII and Y124, respectively, with ± 5 K uncertainty (see arrows in Fig. 3).

Three groups have previously detected this drop in low-field measurements of underdoped Y123, with $B < 15$ T, on crystals with $T_c(0) = 60$-70 K (refs 13,14,15). Because these earlier studies were limited to high temperatures ($T > T_c(0)$), they failed to reveal that the drop is just the start of a large swing to negative values. By measuring $R_{xx}$ and $R_{xy}$ along both $a$ and $b$ axes, Segawa & Ando[15] were able to show that the drop in $R_H(T)$ is a property of the planes, not the chains. From the perfect linearity of $R_{xy}$ vs $B$ they also concluded that the drop is not due to flux flow[15].

The most natural explanation for the negative $R_H$ is the presence of an electron pocket in the Fermi surface. (In principle, it could also come from a hole pocket with portions of negative curvature[16].) In a scenario where the Fermi surface contains both electron and hole pockets, the sign of $R_H$ depends on the relative magnitude of the respective densities, $n_e$ and $n_h$, and mobilities[17], $\mu_e$ and $\mu_h$. ($\mu \equiv e\,\tau\,/\,m^*$, where $e$ is the electron charge, $1\,/\,\tau$ is the scattering rate and $m^*$ is the effective mass.) Given that these materials are hole-doped, we expect $n_h > n_e$. The fact that $R_H < 0$ at low $T$ therefore implies that $\mu_e > \mu_h$ at low $T$. Given strong inelastic scattering, this inequality can then easily invert at high $T$, offering a straightforward mechanism for the sign change in $R_H$. This happens in simple metals like Al and In (ref. 17) and is typical of compensated metals (with $n_e = n_h$). In high-purity samples of NbSe$_2$, a quasi-2D metal that undergoes a charge density wave transition at $T_{CDW} \approx 30$ K, $R_H(T)$ drops from its positive and flat behaviour above $T_{CDW}$ to eventually become negative below $T_0 \approx 25$ K (ref. 18), as reproduced in Fig. S5. In impure samples, however, $R_H(T)$ remains positive at all $T$ (ref. 18; Fig. S5), showing that the electron/hole balance can depend sensitively on impurity/disorder scattering.

A scenario of electron and hole pockets for YBCO resolves a puzzle in relation to the Shubnikov-de Haas oscillations observed in Y123 ortho-II (ref. 2). It is the apparent violation of the Luttinger sum rule, which states that the total carrier density $n$ must be equal to the total area of the 2D Fermi surface. From the oscillation frequency $F = 530$ T, one gets a carrier density $n_{SdH} = 0.038$ carriers/ planar Cu atom per pocket via $F = n_{SdH} \Phi_0$, where $\Phi_0 = 2.07 \times 10^{-15}$ T $m^2$ is the flux quantum. Under the assumption that the pocket is a hole pocket (of arbitrary curvature) and there is nothing else in the Fermi surface, and assuming that $n$ must be equal to the density of doped holes, $i.e.$, $n = p = 0.10$, the Luttinger sum rule is clearly violated, whether the relevant Brillouin zone includes one or two (or any number) of these pockets[19], $i.e.$, whether $n = n_{SdH} = 0.038$ , or $n = 2 n_{SdH} = 0.076$. If, on the other hand, the Fermi surface contains other sheets (not seen in the SdH oscillations) besides the observed pockets, then the sum rule can easily be satisfied.

The fact that $R_H$ is negative at low $T$ implies that the SdH frequency which was seen in Y123-II (ref. 2) must come from the high-mobility electron pocket, since the amplitude of SdH oscillations depends exponentially on mobility, as $\exp(-\pi / \mu B)$. The hole-like portions of the Fermi surface are either open or have a lower mobility at $T \rightarrow 0$. The largest value of $R_H$ that a single electron pocket of density $n_{SdH} = 0.038$ electrons per unit cell can produce is $R_H^{SdH} = -V_{cell} / e\ n_{SdH} = -29$ mm$^3$/C. Within the uncertainty in the geometric factor, this is the magnitude of $R_H$ measured in II and VIII at low temperature (see Fig. 3). (A similar estimate cannot be made for Y124, for in this particular case, unlike in II and VIII with their imperfect CuO chains that localize charge carriers[2], the chains remain metallic down to low temperature, and thus partially short-circuit the Hall voltage, causing a reduction in Hall resistance by as much as a factor 10 or so (see ref. 15).) The picture that emerges is in sharp contrast with the single Fermi "arc" at ($\pi/2$, $\pi/2$) seen in ARPES studies on other cuprates in zero field (see discussion in refs 1,2). A possible explanation for the discrepancy is that ARPES detects only the whole pocket (one side of it) and SdH only the electron pocket (thanks to its high mobility in the elastic scattering regime). This would suggest a Fermi surface similar to that proposed for electron-doped cuprates (see below) , with a hole pocket at ($\pi/2$, $\pi/2$) and an electron pocket at ($\pi$,0).

One might ask why a negative $R_H$ has not been seen in $Bi_2Sr_{2-x}La_xCuO_{6+\delta}$ (BSLCO) and $La_{2-x}Sr_xCuO_4$ (LSCO), the other hole-doped cuprates to have been measured up to 50 T (refs 20,21). Indeed, although $R_H$ can drop by nearly a factor 3 between 150 K and 1.5 K in BSLCO, in a manner not unlike what is seen here in YBCO, it never becomes negative. A possible explanation is that the negative $R_H$ in the YBCO cuprates is associated with the presence of CuO chains. In Y123-II, however, electrical anisotropy in the *ab*-plane is unity below 100 K (ref. 2), implying that the chain sub-system is non-conducting at low temperatures. In Y124, the double chain unit remains metallic down to low $T$ and will therefore have an associated Hall coefficient. However, in isostructural, non-superconducting Pr124, where only the chains are conducting, $R_H$ is found to be positive at low $T$ (ref. 22). A more likely alternative is that the much stronger disorder scattering characteristic of BSLCO and LSCO compared to Y123 or Y124 suppresses $\mu_e$ more severely than $\mu_h$. In such a case, the electron pocket, although present, only manifests itself in the temperature dependence of $R_H$, not its sign. As mentioned above, this is what happens in $NbSe_2$ : adding impurities eliminates the sign change in $R_H$ (see ref. 18 and Fig. S5).

Because the LDA-calculated band structures of Y123 (see ref. 2 and references therein) and Y124 (see ref. 4 and references therein) do not support electron pockets, we are led to the fundamental question: how do they come about? The combination of a small Fermi surface volume from SdH oscillations and a negative $R_H$ pointing to electron pockets, in both Y123 and Y124, argues strongly for a reconstruction of the LDA Fermi surface. The standard mechanism for such reconstruction is the onset of a density-wave (DW) instability, as encountered in numerous materials. Examples in quasi-2D materials include $NbSe_2$, already mentioned, and the transition-metal oxide $Ca_3Ru_2O_7$ (ref. 23).

In cuprates, at least three mechanisms can be invoked for a reconstruction of the Fermi surface that would result in electron and hole pockets. The first is an antiferromagnetic (AF) phase with a $(\pi,\pi)$ ordering wavevector, which causes the large hole-like FS to reconstruct into a small hole pocket at $(\pi/2,\pi/2)$ and a small electron pocket at $(\pi,0)$. This model was used to explain the sign change in the low-temperature $R_H$ measured in the electron-doped cuprate $Pr_{2-x}Ce_xCuO_{4-\delta}$ (PCCO) (ref. 7), upon crossing a critical concentration $x_c$ close to where long-range AF order ends. In PCCO, $R_H$ is positive at all temperatures at high doping ($x > 0.19$) and negative at low doping ($x < 0.15$), but it changes sign at intermediate dopings, from positive below $T_0 = 30\text{-}40$ K to negative above[6]. The latter behaviour is similar (but opposite in sign) to that of YBCO in the range $0.10 < p < 0.14$, suggesting that the transport properties of both materials should be interpreted in terms of electrons and holes with different, $T$-dependent mobilities. In the case of hole-doped cuprates, however, it is less likely that AF order is the relevant mechanism, since long-range AF order is thought to be confined to lower doping ($p < 0.05$). Nevertheless, the possibility should be investigated, in particular since a large magnetic field may push the phase transition to higher doping[19]. The second scenario is a theoretical phase with $d$-density wave (DDW) order, which would also cause a $(\pi,\pi)$ folding of the Fermi surface and thus could produce electron pockets near $(\pi,0)$ (ref. 24). The third scenario is a density-wave phase akin to that encountered experimentally in LBCO and Nd-doped LSCO. In these systems, a Fermi surface reconstruction, signalled by a precipitous drop in $R_H$ (Fig. 1), coincides with the density-wave transition, observed directly via neutron diffraction. Millis & Norman[25] have recently shown that within mean-field theory a "1/8-stripe" spin/charge DW order of this kind does reconstruct the Fermi surface of a generic hole-doped cuprate in a way that tends to produce an electron pocket.

Although the similarities between the $La_2CuO_4$ and YBCO families are highly suggestive, there are some important differences. First, while there is unambiguous direct evidence for static spin/charge DW order in LBCO and Nd-LSCO from neutron diffraction[8,9], there is none so far in YBCO. This could be because the putative DW phase in YBCO involves fluctuating rather than static order[26], or short-range rather than long-range order[27]. Secondly, the anomalies in the transport properties associated with DW order are sharp in LBCO (ref. 11) and Nd-doped LSCO (ref. 10), while in YBCO the temperature dependence of $R_H$ is smooth. Note, however, that anomalies in the former materials appear to be sharp only when the low-temperature structural transition coincides with the DW transition. When structural and DW transitions do not coincide, as in Eu-doped LSCO (ref. 28), the anomalies also appear to be smooth.

In summary, the normal state of underdoped YBCO is characterised by a negative Hall coefficient, revealing the presence of an electron pocket in the Fermi surface whose mobility in these clean cuprates is high enough to outweigh the contribution from other, hole-like parts of the Fermi surface. This implies that the Shubnikov-de Haas oscillations observed recently in the same materials must come from those electron pockets. It also suggests that the generally positive Hall coefficient seen in other hole-doped cuprates results from electron mobilities that are too low because of stronger disorder scattering. As electron pockets are not supported by the band structure of YBCO, we conclude that they must come from a reconstruction of the Fermi surface, which occurs at a critical doping above $p = 0.14$. In the absence of any direct evidence so far for long-range density-wave order in YBCO, our findings call for diffraction experiments to search for them and theoretical investigations of other, more unconventional scenarios.

**METHODS SUMMARY**

**Samples**. The Y123 samples are fully detwinned crystals of $YBa_2Cu_3O_y$ grown in non-reactive $BaZrO_3$ crucibles from high-purity starting materials (see ref. 2, and references therein). For II (VIII), the oxygen content was set at $y = 6.51$ (6.67) and the dopant oxygen atoms were made to order into an ortho-II (ortho-VIII) superstructure, yielding a superconducting transition temperature $T_c = 57.5$ K (66.0 K). The samples are uncut, unpolished thin platelets, whose transport properties are measured via gold evaporated contacts (resistance $< 1$ $\Omega$), in a six-contact geometry. Typical sample dimensions are 20-50 $\times$ 500-800 $\times$ 500-1000 $\mu m^3$ (thickness $\times$ width $\times$ length). The $YBa_2Cu_4O_8$ crystals were grown by a flux method in $Y_2O_3$ crucibles and an $Ar/O_2$ mixture at 2000 bar, with a partial oxygen pressure of 400 bar (ref. 29).

**Estimates of hole doping**. The hole doping $p$ in Y123 is determined from a relationship between $T_c$ and the $c$-axis lattice constant[30]. For our II (VIII) samples, the measured $T_c$ implies $p = 0.099$ (0.120). We assume the doping in Y124 is the same as in a sample of Y123 with the same $T_c$ of 80 K, namely $p = 0.137 \approx 0.14$.

**Resistance measurements**. Longitudinal ($R_{xx}$) and transverse ($R_{xy}$) resistances are obtained from the voltage drop measured diagonally on either side of the sample width, for a field parallel and anti-parallel to the $c$-axis: $R_{xx} \equiv (V_{up} + V_{down}) / 2I_x$ and $R_{xy} \equiv (V_{up} - V_{down}) / 2I_x$. Measurements on II and Y124 were performed at the LNCMP in Toulouse, in a pulsed resistive magnet up to 61 T. Measurements on II and VIII were performed at the NHMFL in Tallahassee, in a steady hybrid magnet up to 45 T.

[1] Julian, S.R. & Norman, M.R. Local pairs and small surfaces. *Nature* **447**, 537-539 (2007).

[2] Doiron-Leyraud, N. *et al.* Quantum oscillations and the Fermi surface in an underdoped high-$T_c$ superconductor. *Nature* **447**, 565-568 (2007).

[3] Yelland, E.A. *et al.* Quantum oscillations in the underdoped cuprate $YBa_2Cu_4O_8$. arXiv:0707.0057.

[4] Bangura, A.F. *et al.* Shubnikov-de Haas oscillations in $YBa_2Cu_4O_8$. arXiv:0707.4461.

[5] Hussey, N.E. *et al.* Observation of a coherent three-dimensional Fermi surface in a high-transition temperature superconductor. *Nature* **425**, 814-817 (2003).

[6] Li, P., Balakirev, F.F. & Greene, R.L. High-field Hall resistivity and magneto-resistance in electron-doped $Pr_{2-x}Ce_xCuO_{4-\delta}$. *Phys. Rev. Lett.* **99**, 047003 (2007).

[7] Lin, J. & Millis, A.J. Theory of low-temperature Hall effect in electron-doped cuprates. *Phys. Rev. B* **72**, 214506 (2005).

[8] Tranquada, J. M. *et al.* Evidence for stripe correlations of spins and holes in copper oxide superconductors. *Nature* **375**, 561-563 (1995).

[9] Ichikawa, N. *et al*. Local magnetic order vs superconductivity in a layered cuprate. *Phys. Rev. Lett.* **85**, 1738-1741 (2000).

[10] Noda, T., Eisaki, H. & Uchida, S. Evidence for one-dimensional charge transport in $La_{2-y-x}Nd_ySr_xCuO_4$. *Science* **286**, 265-268 (1999).

[11] Adachi, T., Noji, T. & Koike, Y. Crystal growth, transport properties, and crystal structure of the single-crystal $La_{2-x}Ba_xCuO_4$ ($x$=0.11). *Phys. Rev. B* **64**, 144524 (2001).

[12] Harris, J.M. *et al.* Hall angle evidence for the superclean regime in 60-K YBa$_2$Cu$_3$O$_{6+y}$. *Phys. Rev. Lett.* **73**, 1711-1714 (1994).

[13] Ito, T., Takenaka, K. & Uchida, S. Systematic deviation from *T*-linear behavior in the in-plane resistivity of YBa$_2$Cu$_3$O$_{7-y}$ : evidence for dominant spin scattering. *Phys. Rev. Lett.* **70**, 3995-3998 (1993).

[14] Wang, Y. & Ong, N.P. Particle-hole symmetry in the antiferromagnetic state of the cuprates. *Proc. Nat. Acad. Sci.* **98**, 11091-11096 (2001).

[15] Segawa, K. & Ando, Y. Intrinsic Hall response of the CuO$_2$ planes in a chain-plane composite system of YBa$_2$Cu$_3$O$_y$. *Phys. Rev. B* **69**, 104521 (2004).

[16] Ong, N.P. Geometric interpretation of the weak-field Hall conductivity in two-dimensional metals with arbitrary Fermi surface. *Phys. Rev. B* **43**, 193-201 (1991).

[17] Ashcroft, N.W. The reversal of Hall fields in Aluminium and Indium. *Phys. Kondens. Materie* **9**, 45-53 (1969).

[18] Huntley, D.J. & Frindt, R.F. Transport properties of NbSe$_2$. *Can. J. Phys.* **52**, 861-867 (1974).

[19] Chen, W.-Q., Yang, K.-Y, Rice, T.M. & Zhang, F.-C. Quantum oscillations in magnetic-field-induced antiferromagnetic phase of underdoped cuprates : application to ortho-II YBa$_2$Cu$_3$O$_{6.5}$. arXiv:0706.3556.

[20] Balakirev, F.F. *et al.* Signature of optimal doping in Hall-effect measurements on a high-temperature superconductor. *Nature* **424**, 912-915 (2003).

[21] Balakirev, F.F. *et al.* Magneto-transport in LSCO high-$T_c$ superconducting thin films. *New J. of Phys.* **8**, 194 (2006).

[22] Horii, S. *et al*. On the dimensionality of the Cu-O double-chain site of $PrBa_2Cu_4O_8$. *Phys. Rev. B* **66**, 054530 (2002).

[23] Baumberger, F. *et al*. Nested Fermi surface and electronic instability in $Ca_3Ru_2O_7$. *Phys. Rev. Lett.* **96**, 107601 (2006).

[24] Chakravarty, S. *et al*. Sharp signature of a $d_{x2-y2}$ quantum critical point in the Hall coefficient of cuprate superconductors. *Phys. Rev. Lett.* **89**, 277003 (2002).

[25] Millis, A.J. & Norman, M.R. Antiphase stripe order as the origin of electron pockets observed in 1/8-hole-doped cuprates. arXiv:0709.0106.

[26] Kivelson, S.A. *et al*. How to detect fluctuating stripes in the high-temperature superconductors. *Rev. Mod. Phys.* **75**, 1201-1241 (2003).

[27] Kohsaka, Y. *et al*. An intrinsic bond-centered electronic glass with unidirectional domains in underdoped cuprates. *Science* **315**, 1380-1385 (2007).

[28] Hücker, M. *et al*. Consequences of stripe order for the transport properties of rare earth doped $La_{2-x}Sr_xCuO_4$. *J. Phys. Chem. Solids* **59**, 1821-1824 (1998).

[29] Adachi, S. *et al*. Crystal growth of $YBa_2Cu_4O_8$. *Physica* **301C**, 123 (1998).

[30] Liang, R., Bonn, D.A. & Hardy, W.N. Evaluation of $CuO_2$ plane hole doping in $YBa_2Cu_3O_{6+x}$ single crystals. *Phys. Rev. B* **73**, 180505 (2006).

**Supplementary Information** is linked to the online version of the paper at www.nature.com/nature.

**Acknowledgements** We thank K. Behnia, L. Brisson, S. Chakravarty, J.C. Davis, R.L. Greene, S.A. Kivelson, G.G. Lonzarich, M.R. Norman, A.J. Schofield, A.-M.S. Tremblay and D. Vignolle for inspiring discussions, and J. Corbin and M. Nardone for their help with the experiments. We acknowledge support from the Canadian Institute for Advanced Research, the LNCMP and the NHMFL, and funding from NSERC, FQRNT, EPSRC, and a Canada Research Chair. Part of this work was supported by the French ANR IceNET and EuroMagNET.

**Author Contributions** D.L. and N.D.-L. contributed equally to this work.

**Author Information** Reprints and permissions information is available at www.nature.com/reprints. The authors declare no competing financial interests. Correspondence and requests for materials should be addressed to C.P. (proust@lncmp.org) or L.T. (louis.taillefer@physique.usherbrooke.ca).

## Figure 1 | Hall resistance of LBCO and YBCO.

Hall resistance $R_{xy}$ vs $T$, normalised at 60 K, for La$_{2-x}$Ba$_x$CuO$_4$ (LBCO) at $p = 0.11$ ($x = 0.11$; from ref. 11) and YBa$_2$Cu$_3$O$_y$ (YBCO) at $p = 0.10$. Our data on YBCO were obtained on two different Y123 ortho-II samples (with $y = 6.51$), one measured in a continuous temperature sweep at a constant field of 45 T (at the NHMFL; blue curve) and the other measured via field sweeps up to 61 T (at the LNCMP; red circles, taken at 55 T).

**Figure 2 | Hall coefficient vs magnetic field.**

Hall coefficient $R_H \equiv t\, R_{xy} / B$ as a function of magnetic field $B$ at different temperatures as indicated: **a)** Y123 ortho-II ($p = 0.10$); **b)** Y123 ortho-VIII ($p = 0.12$); **c)** Y124 ($p = 0.14$). The arrows in (**b**) indicate the fields $B_s$ and $B_n$ described in the text and defined in the Supplementary Information. The 4.2-K isotherm of Y124 illustrates nicely the basic components of $R_H$ : the flat negative part at high field (above $B_n$) is the normal-state value, while the positive overshoot just above $B_s$ is due to a vortex flux-flow contribution.

**Figure 3 | Normal-state Hall coefficient vs temperature.**

Hall coefficient $R_H$ vs $T$ for Y123 - II, Y123 - VIII and Y124 (data multiplied by 10), at $B = 55$, 45 and 55 T, respectively. $T_0$ is the temperature where $R_H$ changes sign, equal to 30, 70 and 30 ± 2 K, respectively. $T_{max}$ is the temperature at which $R_H$ is maximum, equal to 50, 105 and 60 ± 5 K, respectively. The (black) arrow indicates the value of the Hall coefficient expected for a single electron Fermi pocket of the size imposed by SdH oscillations of frequency $F$, namely $R_H^{SdH} = -\, V_{cell} / e\, n_{SdH}$, where $n_{SdH} = F / \Phi_0$ = 0.038 electrons/unit cell. (The data for Y124 is multiplied by a factor 10 to put it on a scale comparable to II and VIII. The order-of-magnitude reduction of the measured Hall voltage comes in large part from the short-circuiting effect of the CuO chains along the $b$-axis which in this stoichiometric material, unlike in II and VIII, remain highly conductive down to low temperature; see text.)

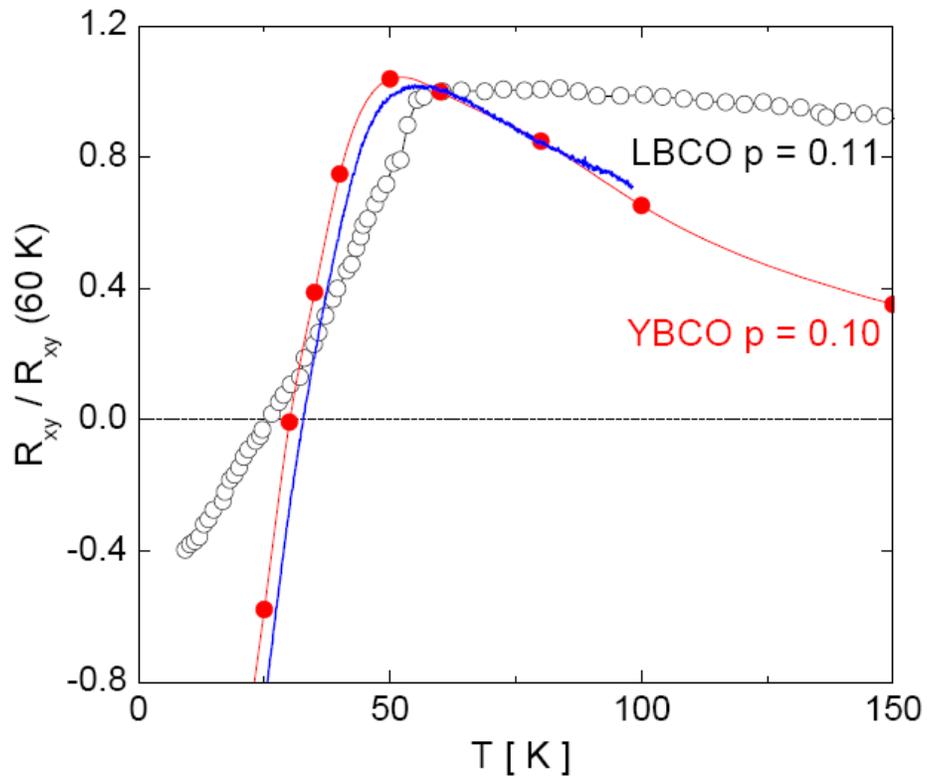

**Figure 1 | Hall resistance of LBCO and YBCO.**

**a**

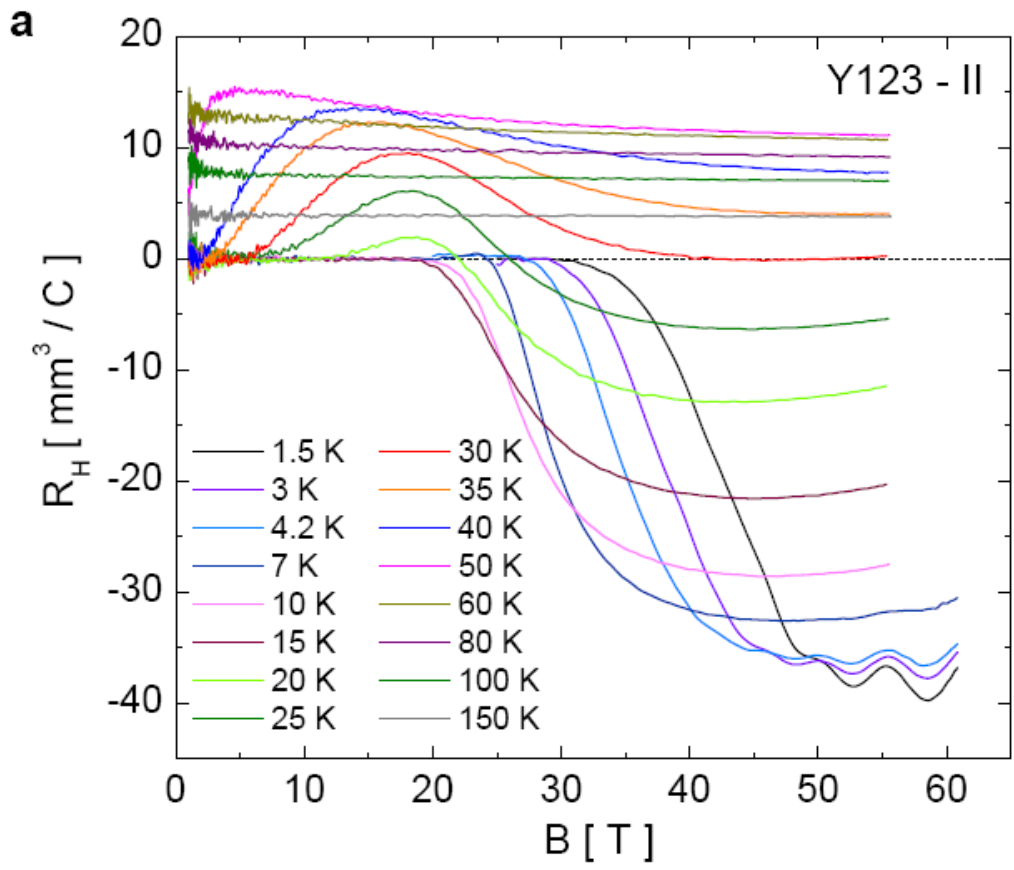

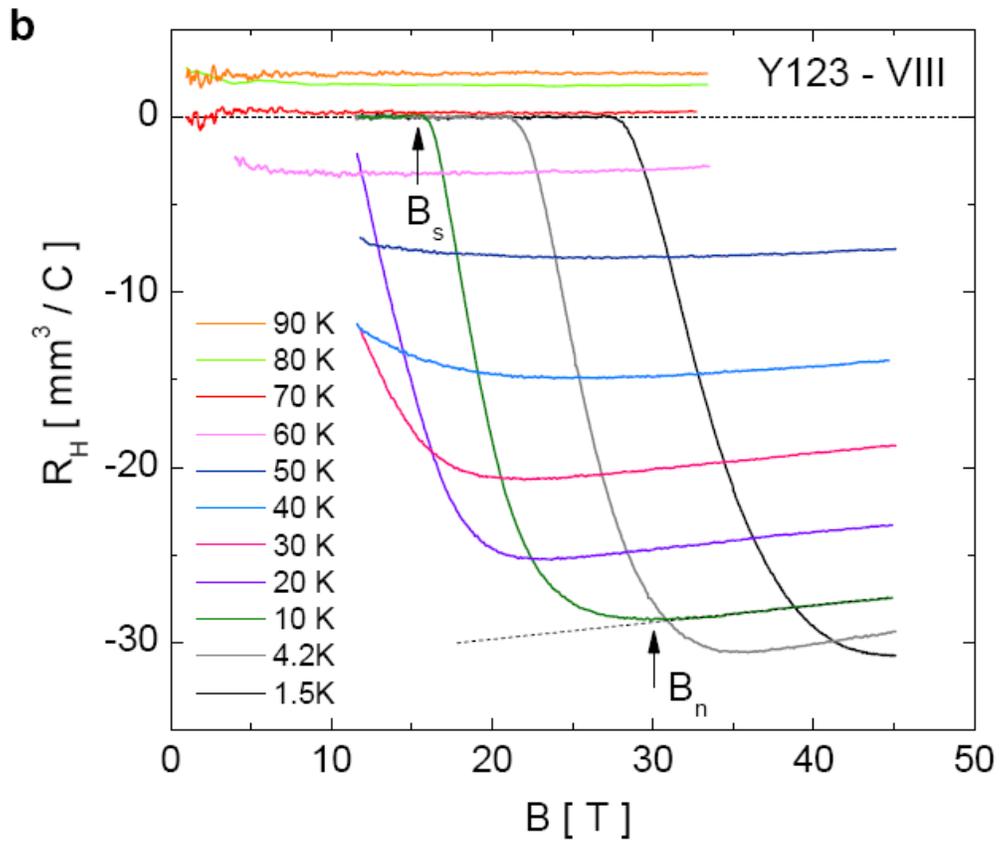

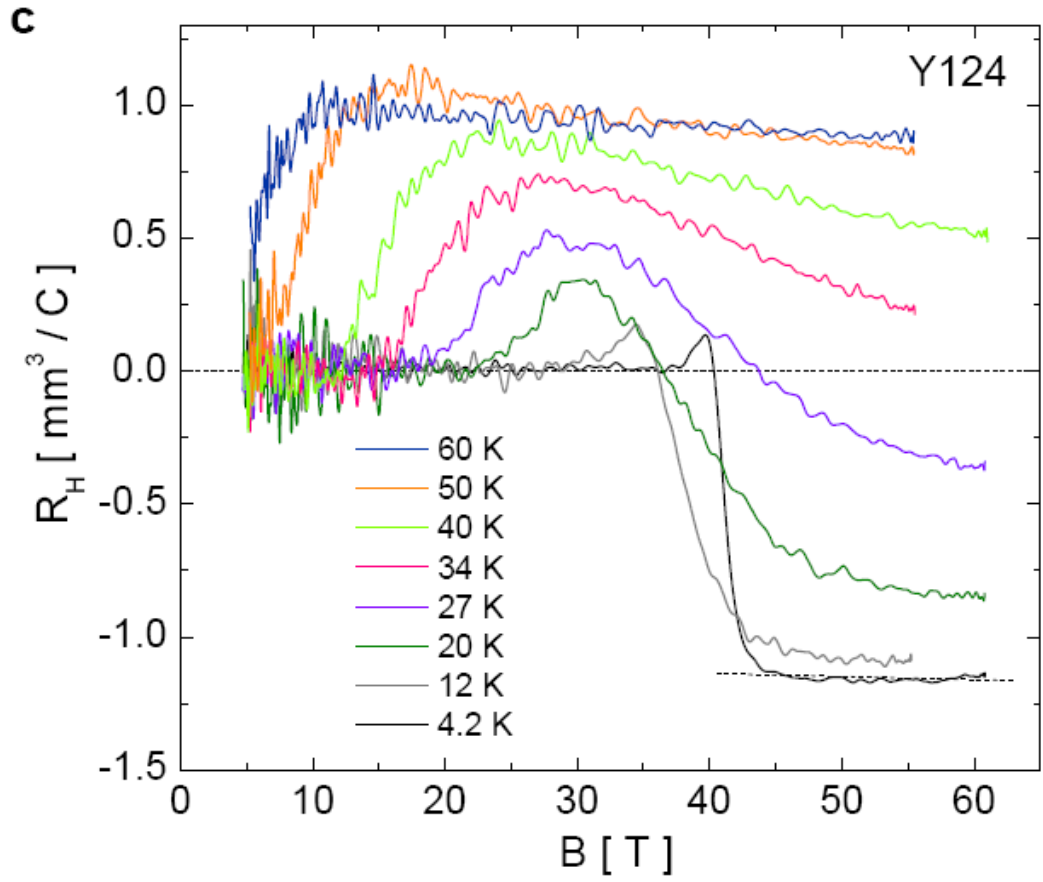

**Figure 2 | Hall coefficient vs magnetic field.**

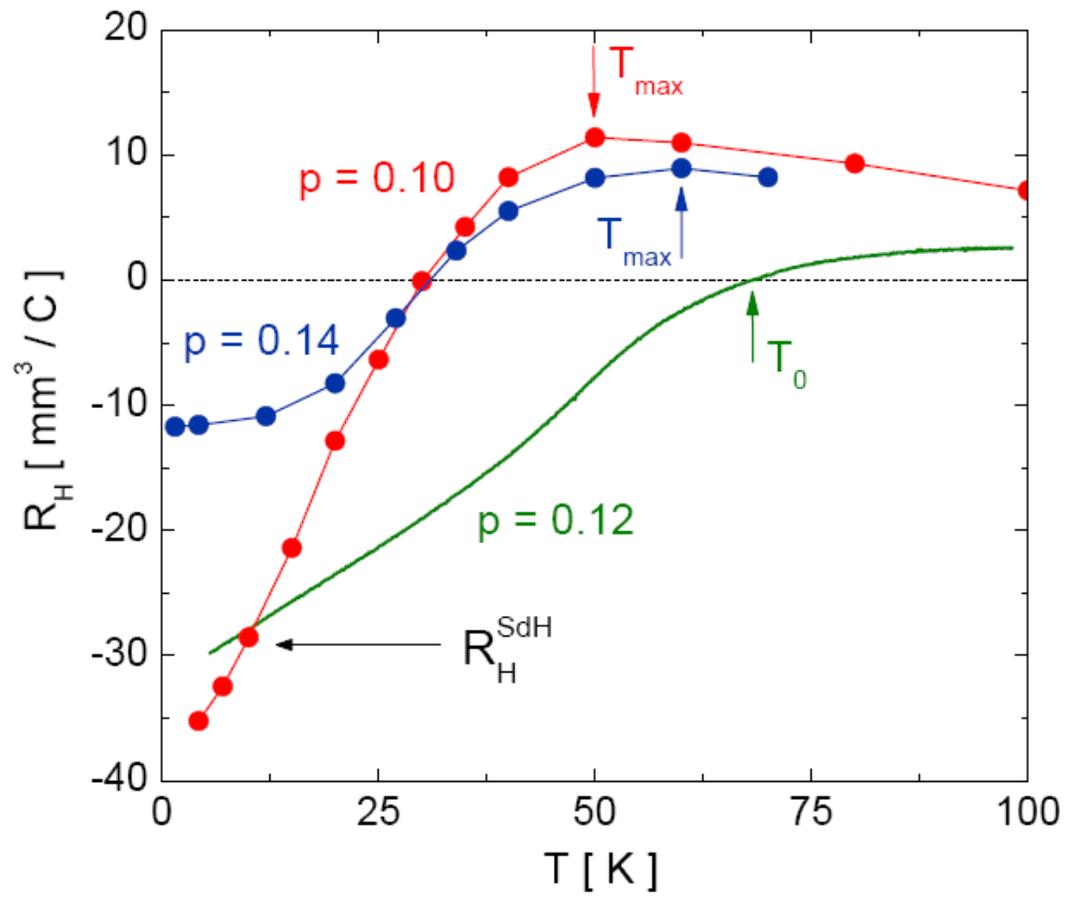

**Figure 3 | Normal-state Hall coefficient vs temperature.**

## Supplementary material for :

## "Electron pockets in the Fermi surface of hole-doped high-$T_c$ superconductors"

David LeBoeuf [1], Nicolas Doiron-Leyraud [1], R. Daou [1], J.-B. Bonnemaison [1],

Julien Levallois [2], N.E. Hussey [3], Cyril Proust [2], L. Balicas [4], B. Ramshaw [5],

Ruixing Liang [5,6], D.A. Bonn [5,6], W.N. Hardy [5,6], S. Adachi [7] &  Louis Taillefer [1,6]

[1] *Département de physique and RQMP, Université de Sherbrooke, Sherbrooke J1K 2R1, Canada*

[2] *Laboratoire National des Champs Magnétiques Pulsés (LNCMP), UMR CNRS-UPS-INSA 5147, Toulouse 31400, France*

[3] *H.H. Wills Physics Laboratory, University of Bristol, Bristol BS8 1TL, UK*

[4] *National High Magnetic Field Laboratory, Florida State University, Tallahassee, Florida 32306, USA*

[5] *Department of Physics and Astronomy, University of British Columbia, Vancouver V6T 1Z4, Canada*

[6] *Canadian Institute for Advanced Research, Toronto M5G 1Z8, Canada*

[7] *Superconductivity Research Laboratory, International Superconductivity Technology Center, Shinonome 1-10-13, Koto-ku, Tokyo 135-0062, Japan*

## Magnetic field – Temperature phase diagrams

In this section, we determine the field scale above which the effects of vortex motion and superconducting fluctuations (of phase or amplitude) have become negligible in $R_{xx}$ and $R_{xy}$, such that these transport coefficients reflect predominantly the properties of the normal state. This field scale, which we label $B_n(T)$ and define below, is plotted in a $B$-$T$ diagram for each of the three materials in Figs. S2a to S2c. It is compared to two other curves. The first is $B_S(T)$, the field below which the vortex solid phase exists, *i.e.* the so-called irreversibility field below which $R_{xy}(B, T) = R_{xx}(B, T) = 0$. It is straightforward to obtain $B_S$ from the isotherms of $R_H(B)$ in Fig. 2. In Fig. S3a, the isotherms at $T = 4.2$ K are shown to respectively yield $B_S = 25$, 20 and 37 T, for II, VIII and Y124, with ± 2 T uncertainty. The second curve to which we compare $B_n$ is $T_0(B)$, defined as the temperature at which $R_H(T)$ changes sign, *i.e.* where $R_H(T_0) = 0$, readily obtained from the curves of $R_H$ vs $T$ in Fig. S1. *The key observation is that $T_0$ is independent of field at the highest fields in all three materials*. This shows that the temperature-induced sign change in $R_H$ at high fields, featured in Fig. 3, is not caused by flux flow (see below). The case is particularly clear in VIII, where the sign change occurs *above* $T_c(0)$ and is totally independent of field over the entire field range from 0 to 45 T (see Fig. S2b). The negative $R_H$ is thus clearly a property of the normal state, the consequence of a drop in $R_H(T)$ which starts below a field-independent temperature $T_{max}$. The value of $T_{max}$ at the three doping levels studied here is 50, 105 and 60 K, for II, VIII and Y124, respectively, with ± 5 K uncertainty.

## Vortex contribution to Hall resistance

It is important to distinguish the field-independent, *temperature-induced* sign change mentioned above, which persists to the highest fields, from the temperature-dependent, *field-induced* sign change that can result from a vortex contribution to $\sigma_{xy}$.

The latter has been studied extensively in a number of cuprates[1],[2] and it arises from a cancellation of normal-state ($\sigma^n_{xy}$) and flux-flow ($\sigma^f_{xy}$) contributions in the total Hall conductivity ($\sigma_{xy} = \sigma^n_{xy} + \sigma^f_{xy}$) that can occur if the two contributions happen to be of opposite sign. Given the very different field dependencies of the two contributions, a cancellation and sign change in $\sigma_{xy}$ (and hence in $R_{xy}$) can only occur at a particular value of $B$, for any given $T$. The most compelling argument against this being the mechanism for the high-field sign change in YBCO is provided by the 70-K isotherm in VIII (see Fig. 2b), which is totally flat and nearly zero at all $B$, *i.e.* $R_H(70 \text{ K}) \approx 0$ independent of $B$. It is indeed unphysical to suppose that a finite $\sigma^f_{xy}$ conspires to remain equal to $-\sigma^n_{xy}$ at all $B$.

A vortex contribution is also observed in our samples, and it makes a *positive* contribution to $R_{xy}$ in all three materials. It is clearly seen in the 4.2 K isotherm of Y124, for example, where it shows up as a small positive overshoot just above $B_S = 37$ T (see Fig. 2c). This vortex-related contribution persists to high temperature, causing a detectable "bump" at low fields up to 50 K or so. In Y124, the low-field regime (in this case below 40 T or so) is dominated by the vortex contribution so that one needs to go above 50 T to uncover the clean normal-state behaviour. The same field-induced positive overshoot is seen in II (Fig. 2a), although not below 20 K, presumably because of the stronger pinning in this non-stoichiometric material. It is instructive to also look at $R_H$ vs $T$, as in Fig. S1a, for we can see then that the 15 T curve lies *above* the 55 T curve at all $T$ up to 80 K (see Fig. S4a). This excess in $R_H(T)$ relative to the normal state curve, *i.e.*, the difference between the two curves, $R_H(T, 15 \text{ T}) - R_H(T, 55 \text{ T})$, is the positive vortex contribution, plotted in the inset of Fig. S4a. It persists at temperatures slightly above $T_c(0)$, as one might expect from superconducting fluctuations. In Fig. S4b, this excess is shown to track the reported rise in the Nernst signal, believed to be a measure of vortex fluctuations[3], in a Y123 sample with the same $T_c$ of 57 K (ref. 4). As in all previous studies (see refs. 1 and 2, and references therein), the vortex contribution to $R_{xy}$ is seen to vanish at high fields (see Fig. S1), specifically above 35, 25 and 50 T in II, VIII and Y124, respectively. Note, however, that the effect of vortices appears to remain detectable up to higher fields in the Nernst signal[3] than it does in the Hall signal.

Flux flow does not cause a sign change in VIII. Looking at the *B-T* diagrams of Fig. S2, one can see that this is because the $T_0(B)$ line lies completely above the superconducting state in the case of VIII, whereas it intersects the $B_n(T)$ line in the other two materials. The bending of $T_0(B)$ that results from this intersection is what gives the field-induced sign change. In general, flux flow appears to have little impact on $R_H$ in this material.

**Magnetic field scales**

As in previous high-field studies of cuprates (refs 5, 6), the fields $B_s$ and $B_n$ are respectively defined at low temperature as the field above which $R_{xy}(B)$ departs from zero and the field below which $R_{xy}(B)$ departs from its high-field, roughly linear behaviour. This is illustrated (by arrows) in Fig. S3a for $T = 4.2$ K. Note that this criterion does not work at high temperatures, so instead we use $R_{xx}(B)$ to determine $B_s$ and $B_n$. At low temperatures, the values obtained by both the longitudinal and transverse resistive components correlate well, as shown in Fig. S3b, giving us confidence in their determination.

[1] Hagen, S.J. *et al.* Anomalous flux-flow Hall effect: $Nd_{1.85}Ce_{0.15}CuO_{4-y}$ and evidence for vortex dynamics. *Phys. Rev. B* **47**, 1064-1068 (1993).

[2] Nagaoka, T. *et al.* Hall anomaly in the superconducting state of high-$T_c$ cuprates: universality in doping dependence. *Phys. Rev. Lett.* **80**, 3594-3597 (1998).

[3] Wang, Y. *et al.* High field phase diagram of cuprates derived from the Nernst effect. *Phys. Rev. Lett.* **88**, 257003 (2002).

[4] Rullier-Albenque, F. *et al.* Nernst effect and disorder in the normal state of high-$T_c$ cuprates. *Phys. Rev. Lett.* **96**, 067002 (2006).

[5] Balakirev, F.F. *et al.* Signature of optimal doping in Hall-effect measurements on a high-temperature superconductor. *Nature* **424**, 912-915 (2003).

[6] Balakirev, F.F. *et al.* Magneto-transport in LSCO high-$T_c$ superconducting thin films. *New J. of Phys.* **8**, 194 (2006).

[7] Huntley, D.J. & Frindt, R.F. Transport properties of $NbSe_2$. *Can. J. Phys.* **52**, 861-867 (1974).

**Figure S1 | Hall coefficient vs temperature.**

Hall coefficient $R_H = t\, R_{xy}\, /\, B$ as a function of temperature $T$ at different fields as indicated for **a)** Y123 ortho-II ($y = 6.51$; $p = 0.10$); **b)** Y123 ortho-VIII ($y = 6.67$; $p = 0.12$); **c)** Y124 ($p = 0.14$).

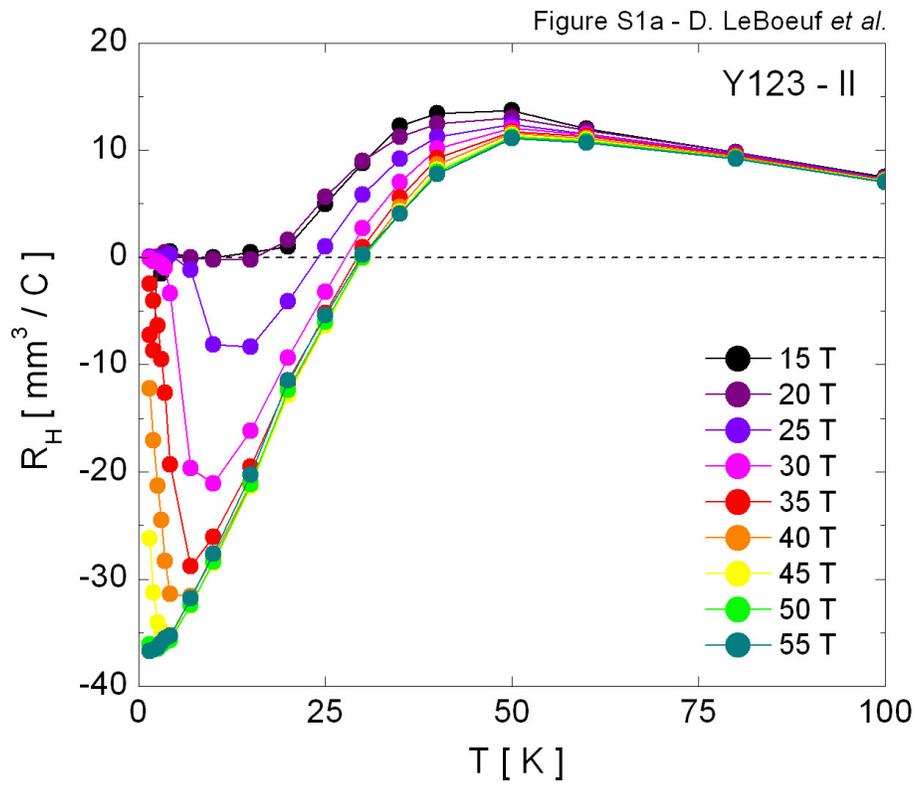

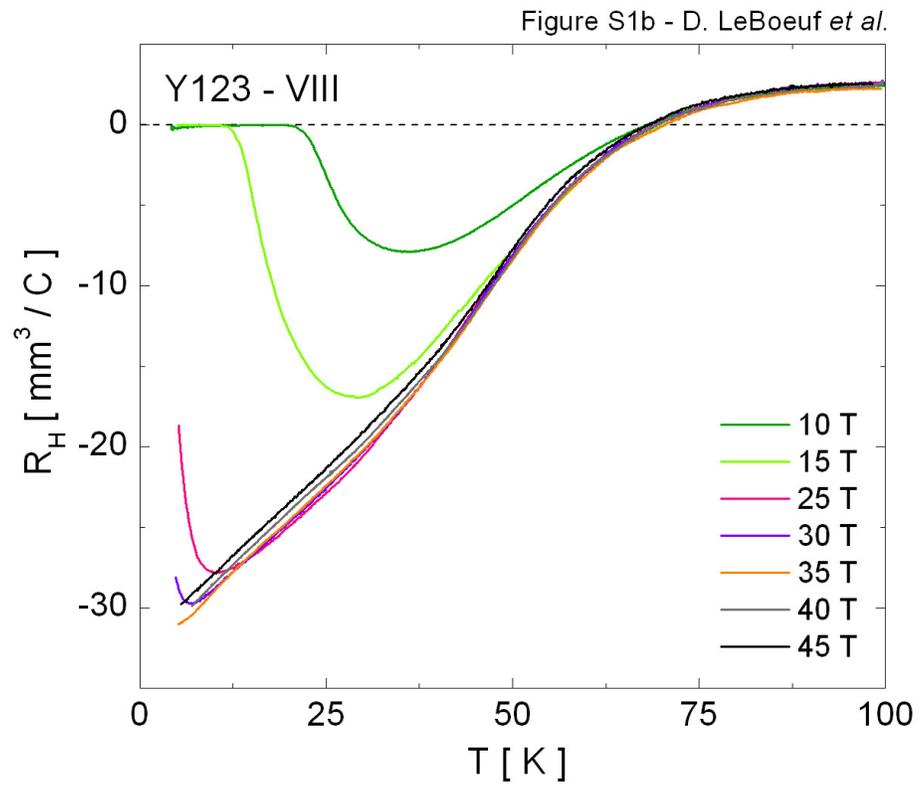

Figure S1b - D. LeBoeuf *et al.*

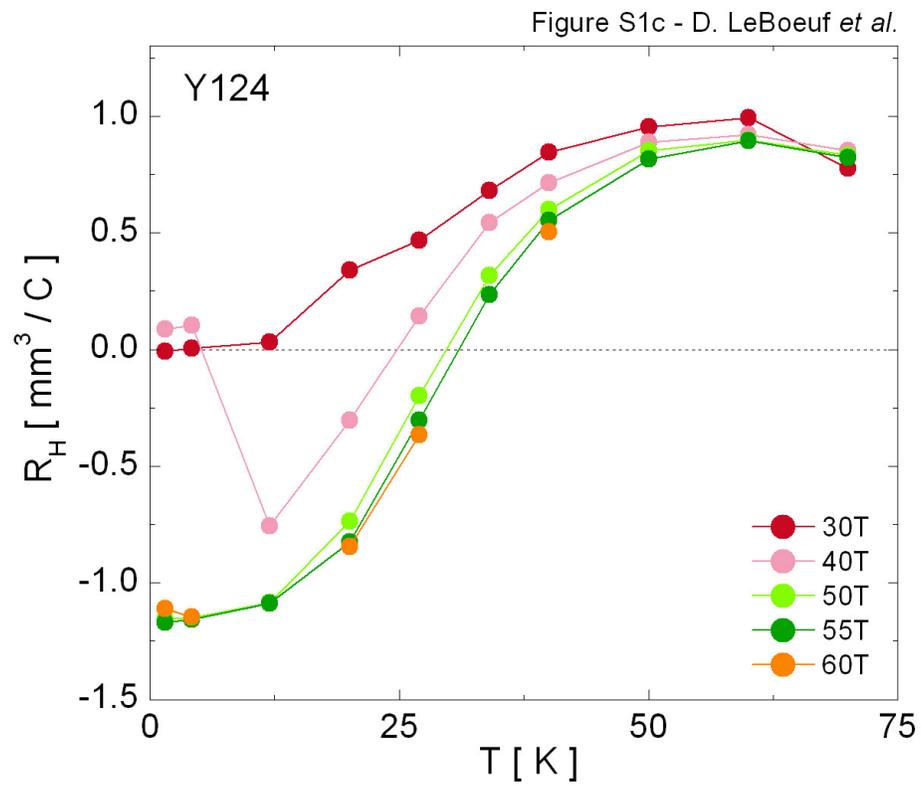

**Figure S2 | Magnetic field - Temperature phase diagram.**

$B - T$ phase diagram for: **a)** Y123 ortho-II ($p = 0.10$); **b)** Y123 ortho-VIII ($p = 0.12$); **c)** Y124 ($p = 0.14$). The vortex solid phase ends at $B_s(T)$ and the transport properties of the normal state are reached above $B_n(T)$, where vortex contributions to transport are negligible. $B_s(T)$ and $B_n(T)$ are defined via $R_{xx}$ (circles) or $R_H$ (squares) vs $B$, as described in the text and shown in Figs. S3a and S3b. Red circles mark $T_0(B)$, the temperature where $R_H$ changes sign.

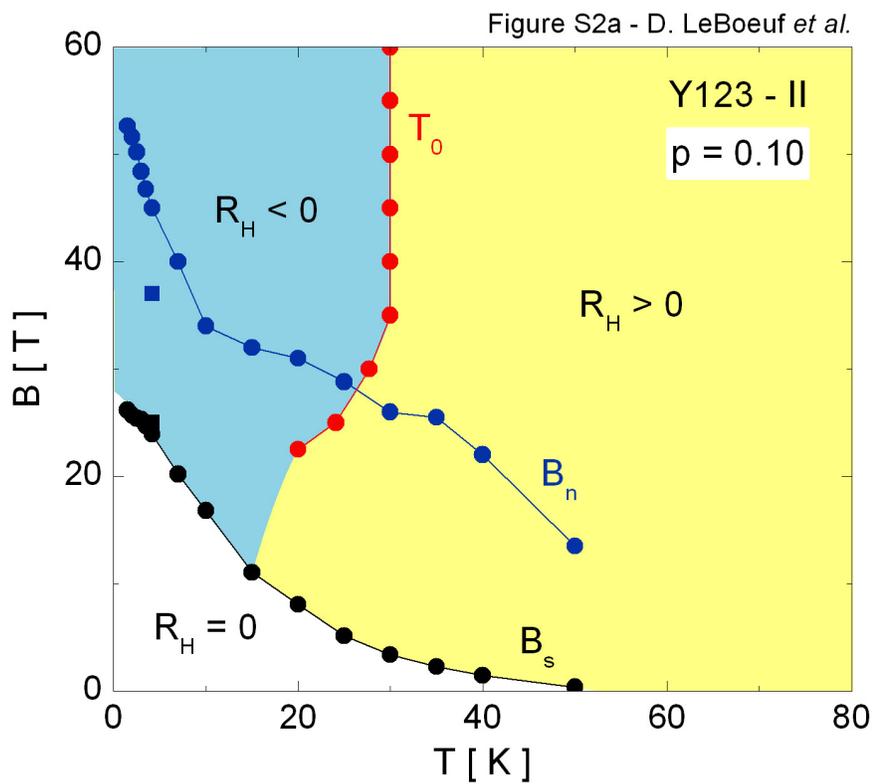

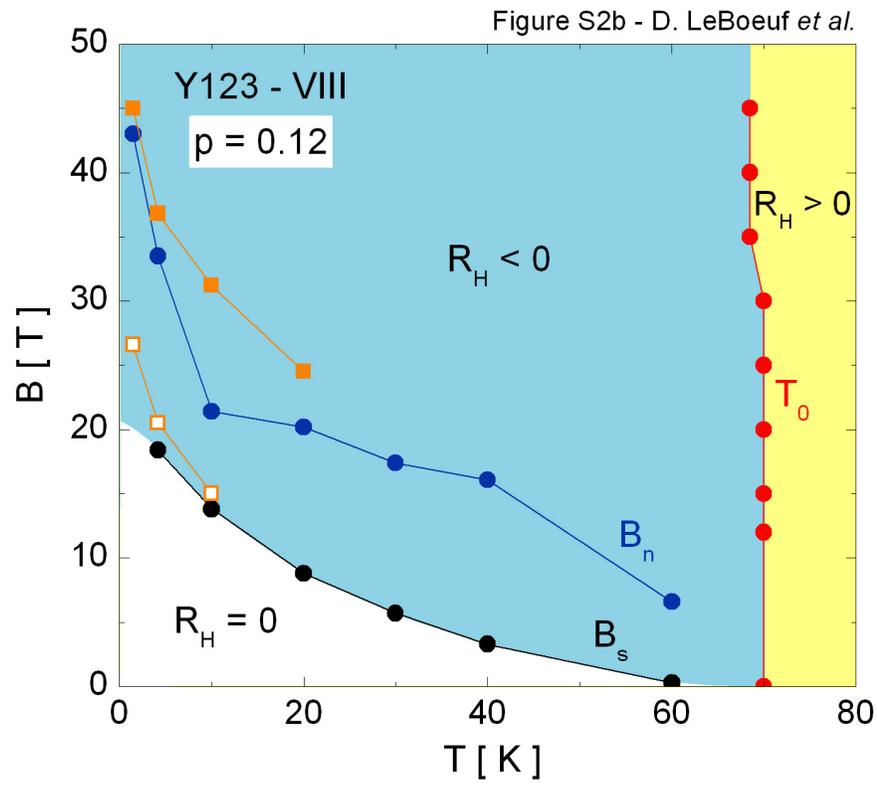

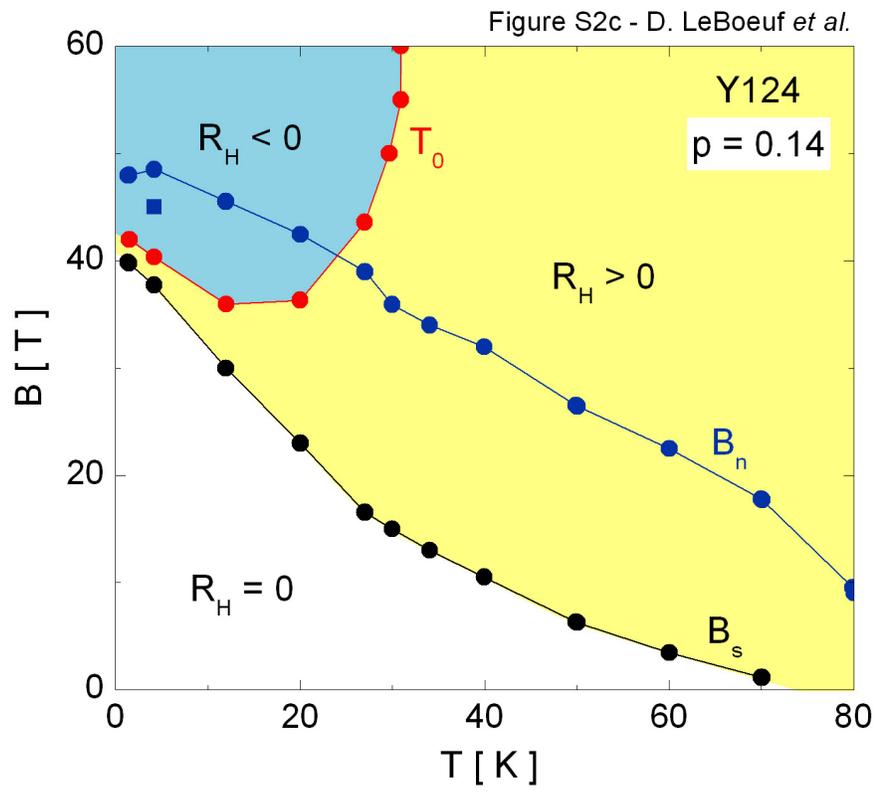

Figure S2c - D. LeBoeuf *et al.*

**Figure S3 | Determination of magnetic field scales.**

**a)** Hall coefficient $R_H = t\,R_{xy}\,/\,B$ as a function of magnetic field $B$ at 4.2 K for Y123 ortho-II ($p = 0.10$), Y123 ortho-VIII ($p = 0.12$), and Y124 ($p = 0.14$). The arrows mark the field $B_s$ above which $R_H$ departs from zero and the field $B_n$ below which $R_H$ deviates from the high-field behaviour of a nearly flat $R_H$ (the fact that the normal-state $R_H$ in II and VIII shows some field dependence is consistent with the two-carrier picture discussed in the text). The values of $B_s$ and $B_n$ thus obtained are shown as squares in Figs. S2a to S2c. (Note that the data shown here for II comes from a different sample to that shown in Fig. 2a, one in which the transition at $B_n$ is sharper and thus more easily defined.) **b)** Longitudinal resistance $R_{xx}$ as a function of field at different temperatures for Y123 ortho-VIII (offset for clarity). The arrows indicate the values for $B_n$ obtained from $R_H$ in a) and in Fig. 2b, showing a good correlation with the field below which $R_{xx}$ departs from its high-field behaviour of a roughly linear magneto-resistance (fitted to a dashed line). The values of $B_n$ thus obtained from $R_{xx}$ are shown as circles in Figs. S2a to S2c.

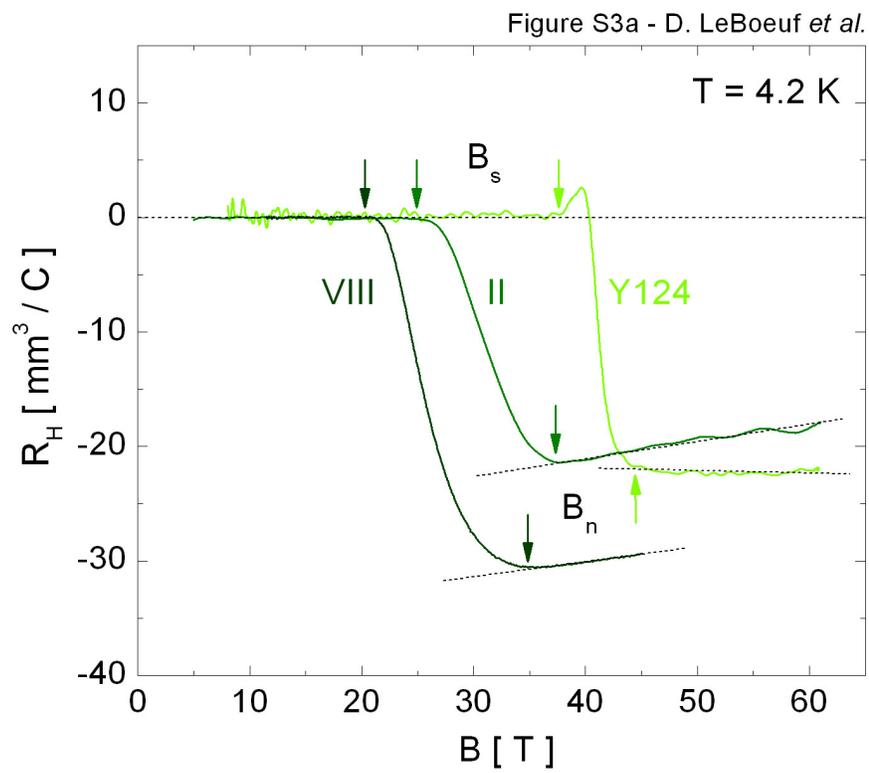

Figure S3a - D. LeBoeuf *et al.*

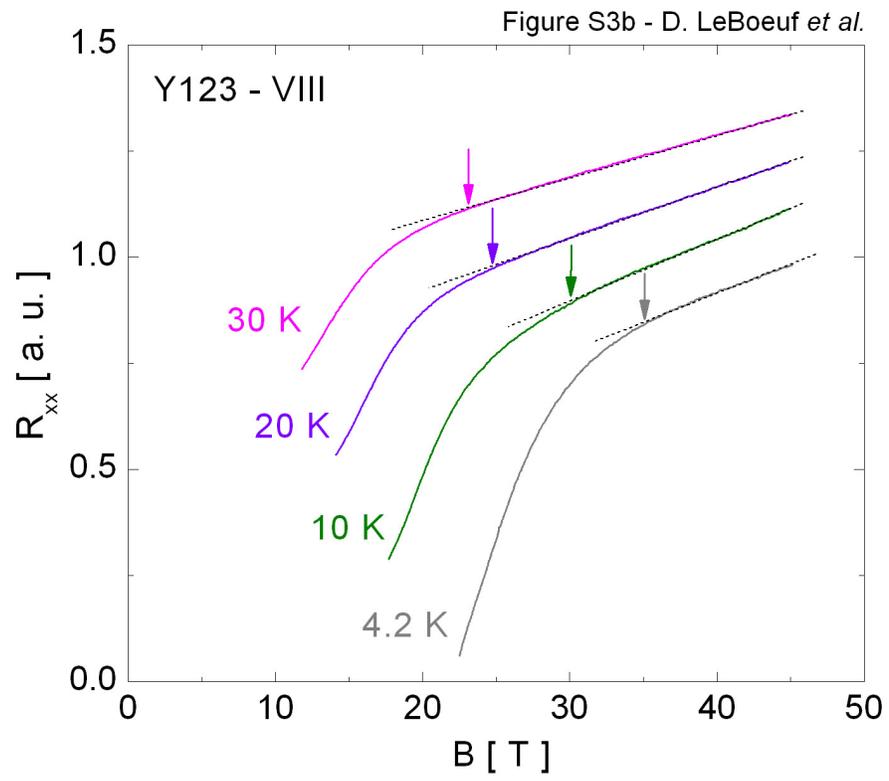

Figure S3b - D. LeBoeuf *et al.*

**Figure S4 |  Vortex contribution to Hall coefficient.**

**a)** Hall coefficient $R_H = t\,R_{xy}\,/\,B$ as a function of temperature $T$ at 15 and 55 T for Y123 ortho-II. Inset: Difference $\Delta R_H = R_H(15T) - R_H(55T)$ between the two curves shown in the main panel. **b)** Difference $\Delta R_H = R_H(15T) - R_H(55T)$ (left scale) and Nernst signal at 8 T on Y123 with $p = 0.10$ (right scale, from ref. 4) as a function of temperature, showing that the positive difference between $R_H(15T)$ and $R_H(55T)$ is caused by a vortex (flux-flow) contribution to the Hall coefficient at low field.

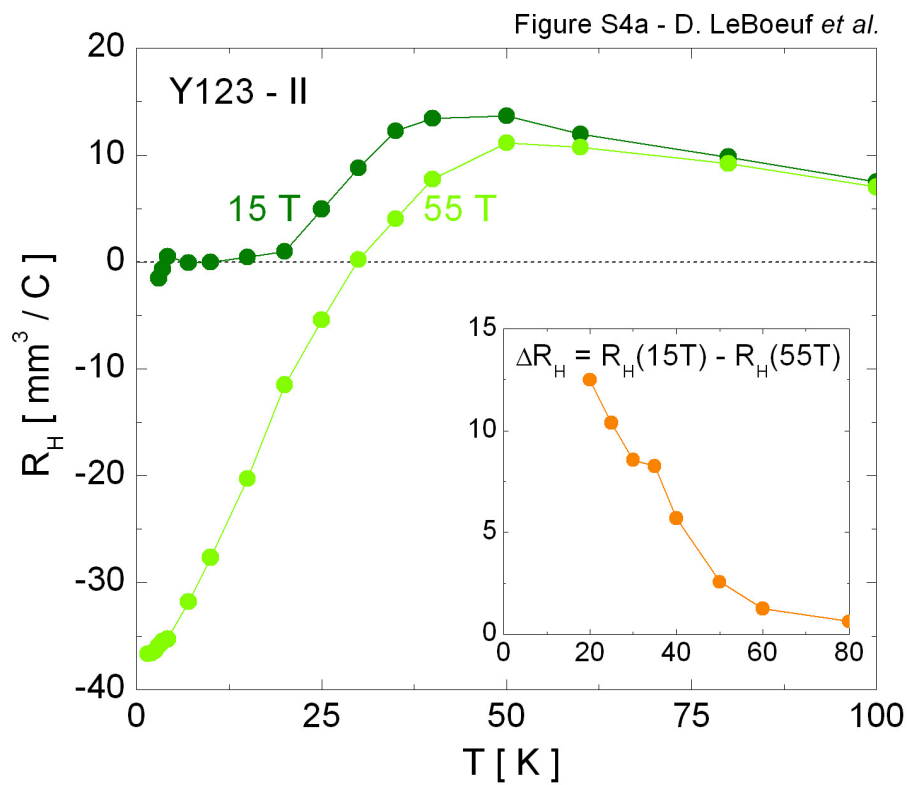

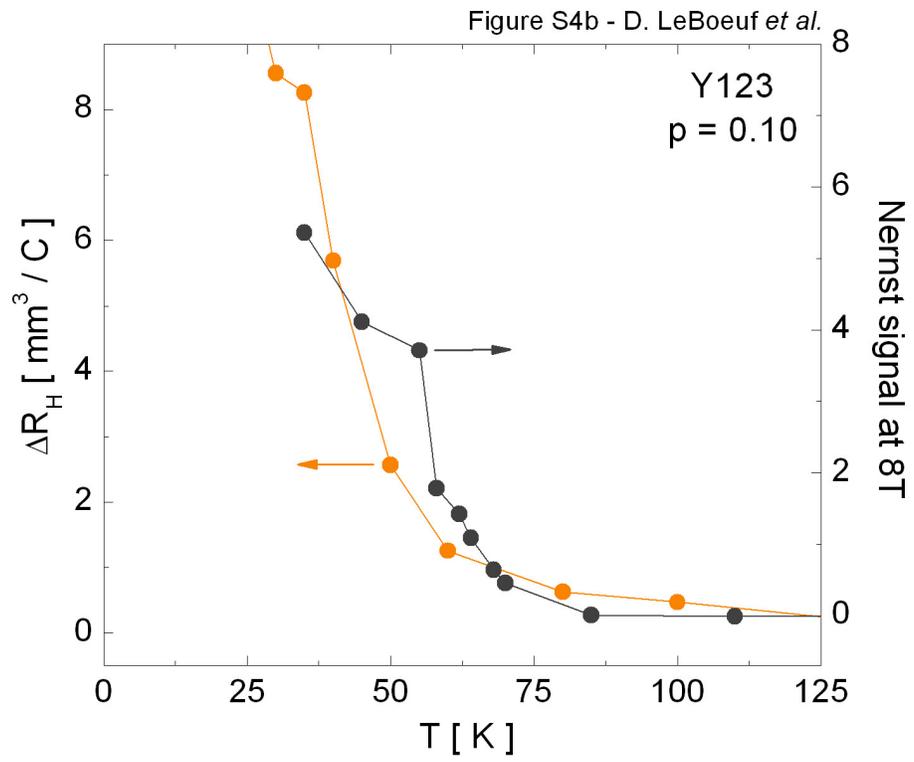

Y123
p = 0.10

**Figure S5 | Hall resistance in NbSe₂.**

Hall resistance $R_{xy}$ normalised at 60 K as a function of temperature for a pure and a dirty sample of NbSe₂ (from ref. 7: samples Q and D, respectively). The vertical dashed line marks the transition to the charge-density-wave phase in NbSe₂ at $T_{CDW} \approx 30$ K.

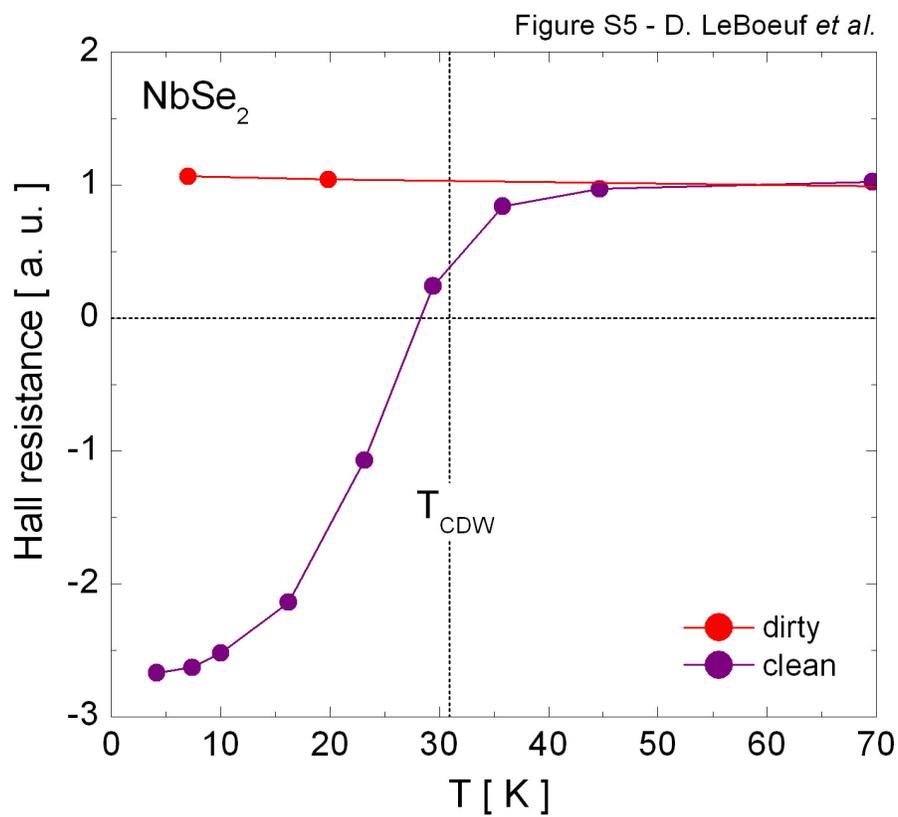